\documentclass{amsart}
\usepackage{graphicx}
\usepackage{amscd}

\newtheorem{theorem}{Theorem}
\theoremstyle{plain}

\newtheorem{corollary}{Corollary}

\newtheorem{lemma}{Lemma}

\newtheorem{proposition}{Proposition}
\newtheorem{remark}{Remark}

\numberwithin{equation}{section}

\begin{document}

\title[Increments of k-spacings]{Strong limits related to the oscillation modulus of the empirical process based on the k-spacing process}

\author{Gane Samb LO}

.

\begin{abstract}
\large
Recently, several strong limit theorems for the oscillation moduli of the
empirical process have been  given in the iid-case. We show that,
with very slight differences, those strong results are also obtained for
some representation of the reduced empirical process based on the
(non-overlapping) k-spacings generated by a sequence of independent random
variables (rv's) uniformly distributed on $(0,1)$. This yields weak limits
for the mentioned process. Our study includes the case where the step k is
unbounded. The results are mainly derived from several properties concerning
the increments of gamma functions with parameters k and one.
\end{abstract}

\maketitle

\Large

\noindent \textbf{Nota-Bene.} This paper was part of the PhD thesis at Cheikh Anta Diop University, 1991, not yet published in a peer-reviewed journal by August 2014. A slightly different version was published in \textit{Rapports Techniques, LSTA, Universit\'e Paris VI}, 48, 1986, under the same title.\\

\section{Introduction and statement of the results} \label{sec1}

\noindent Consider $U_{1},...,U_{n}$ a sequence of independent rv's uniformly
distributed on $(0,1)$, and let 
\begin{equation*}
U_{0,n}=0\leq U_{1,n}\leq ...\leq U_{n,n}\leq U_{n+1,n}=1
\end{equation*}
\noindent be their order statistics. The rv's 
\begin{equation*}
D_{i,n}^{k}=U_{ki,n}-U_{\left( i-1\right) k,n},1\leq i\leq \left[ \frac{n+1}{%
k}\right] =N,
\end{equation*}
\noindent where $\left[ x\right] $ denotes the integer part of $x$, are called the
non-overlapping $k$-spacings. Throughout, we shall assume that $N$ and $k$ are
given and that $n$ is defined by $n=\inf \left\{ j,\left[ \frac{j+1}{k}\right]
=N\right\} $ and then we will be able to study all our sequences as indexed
by $N$ since $k$ will be either fixed or function of $N$.\\

\noindent The study of the properties of $D_{i,n}^{k}$ was introduced by Pyke \cite{8}
and several related papers have appeared in recent years (see e.g. \cite{3}%
). One of the problem concerning the k-spacings is the study of the
empirical process associated with $Nk$ $D_{i,n}^{k},1\leq i\leq N$.\\

\noindent In order to give a comprehensible definition of that process, we recall the
following representation which can be found in \cite{1} in the case where $(n+1)/k$ is an integer :

\begin{equation}
\left\{ D_{i,n}^{k},1\leq i\leq N\right\} =^{d}\left\{ \frac{Y_{i}}{S_{n+1}}%
,1\leq i\leq N\right\} =:\left\{ \frac{\left( \sum_{j=\left( i-1\right)
k+1}^{j=ik}E_{i}\right) }{S_{n+1},},1\leq i\leq N\right\} ,  \label{1.1}
\end{equation}

\noindent where $=^{d}$ denotes the equality in distribution and $S_{n}$ is the
partial sum associated with $E_{1},...,E_{n}$, a sequence of independent and
exponential rv's with meanone, i.e., $S_{n}=E_{1}+...+E_{n}$. Thus, it
follows that, if $\frac{\left( n+1\right) }{k}$ is an integer, the limiting
distribution function of Nk $D_{i,n}^{k},$ for any i and k fixed, is

\begin{equation*}
H_{k}\left( x\right) =\int_{0}^{x}\frac{t^{k-1}e^{-t}}{\left( k-1\right) !}%
dt,\text{ }x\geq 0.
\end{equation*}

\noindent Therefore the empirical process (E.P.) associated with $NkD_{i,n}^{k},1\leq
i\leq N$, may be defined by

\begin{equation}
\beta _{N}\left( x\right) =N^{\frac{1}{2}}\left\{ F_{N}\left( x\right)
-H_{k}\left( x\right) \right\} ,0\leq x\leq +\infty ,  \label{1.2}
\end{equation}

\noindent where $F_{N}$ is the empirical distribution (E.D.F.) of $Nk$ $%
D_{i,n}^{k},1\leq i\leq N$, with

\begin{equation}
F_{N}\left( x\right) =\#\frac{\left\{ i,1\leq i\leq N,\text{ }Nk\text{ }%
D_{i,n}^{k}\leq x\right\} }{N},x\geq 0.  \label{1.3}
\end{equation}

\noindent Straightforward manipulations from (\ref{1.1}), (\ref{1.2}) and (\ref{1.3})
as given in \cite{1} show that even in the general case where $\left(
N-1\right) k\leq n+1\leq Nk$, the reduced process $\alpha _{N}\left(
s\right) =\beta _{N}\left( H_{k}^{-1}\left( s\right) \right) ,0\leq s\leq 1$%
, satisfies

\begin{equation}
\left\{ \alpha _{N}\left( s\right) ,0\leq s<1\right\} =^{d}\left\{ N^{\frac{1%
}{2}}\left\{ \xi _{N}\left( \delta _{n}H_{k}^{-1}\left( s\right) \right)
-s\right\} +0\left( N^{-\frac{1}{2}}\right) ,0\leq s\leq 1\right\} ,
\label{1.4}
\end{equation}

\noindent where $H_{k}^{-1}$ is the inverse function of $H_{k},$ $\xi _{N}$ is the
E.D.F. pertaining to $Y_{1},...,Y_{n}$ and $\delta _{n}=\frac{S_{n+1}}{Nk}.$\\

\noindent The aim of this paper is to give the behavior of the oscillation modulus of $%
\alpha _{N}\left( .\right) $ both where $k$ is fixed and where $k\uparrow
+\infty $. To this end we define 
\begin{equation*}
\wedge _{N}\left( a_{N},R_{N}\right) =\sup_{0\leq h\leq a_{N}}\sup_{0\leq
s\leq 1-h}\left| R_{N}\left( s+h\right) -R_{N}\left( s\right) \right|
\end{equation*}
and 
\begin{equation*}
k_{N}\left( a_{N},R_{N}\right) =\frac{\wedge _{N}\left( a_{N},R_{N}\right) }{%
\left( 2a_{N}\log \log a_{N}^{-1}\right) ^{\frac{1}{2}}},
\end{equation*}
for any sequence of functions $R_{N}\left( s\right) ,0\leq s\leq 1$ and for
any sequence $\left( a_{N}\right) _{N\geq 1},0<a_{N}<1.$ The properties of $%
\wedge _{N}\left( a_{N},R_{N}\right) $, the oscillation modulus of $R_{N}$,
have been first described by Cs\"{o}rgo and R\'{e}v\`{e}sz \cite{2} and
Stute \cite{10} when $R_{N}$ represents the E.P. pertaining to a sequence of
independent and uniformly distributed rv's with 
\begin{equation}
Na_{N}\rightarrow +\infty, \tag{S1}
\end{equation}

\begin{equation}
\frac{\left( \log
a_{N}^{-1}\right) }{\left( Na_{N}\right) }\rightarrow 0 \tag{S2}
\end{equation}

\noindent and 

\begin{equation}
\frac{\left( \log a_{N}^{-1}\right) }{\log \log N\rightarrow +\infty } \tag{S3}
\end{equation}

\noindent  as $N\rightarrow +\infty$.\\

\noindent Later, Mason, Shorack and Wellner (MSW) \cite{7} dealt with
the same for several choices of $\left( a_{N}\right) $ and give among the
results an Erd\"{o}s-Renyi law.\\

\noindent The chief achievement of this paper is the extension of those limit results
to some sequence of process $\bar{\alpha}_{N}$ equal in distribution to $%
\alpha _{N}$. In fact, the fundemental role is played here by the properties
of the tails of the gamma function $H_{k}^{^{\prime }}\left( .\right) $, the
derivative funtion of $H_{k}$. These properties are established in Section
2\ through technical lemmas and the proofs of the following results are
given in Section 3.\\

\bigskip

\begin{theorem} \label{tA}
Let k be fixed. Then, there exists a sequence of processes $\alpha
_{N}^{-}\left( s\right) ,0\leq s\leq 1,$ $N=1,2,...$ such that

\begin{equation}
\forall N\geq 1,\left\{ \alpha _{N}\left( s\right) ,0\leq s\leq 1\right\}
d_{=}\left\{ \alpha _{N}^{-}\left( s\right) ,0\leq s\leq 1\right\} .
\end{equation}

\noindent (I) If $\left( a_{N}\right) _{N\geq 1}$ is a sequence of non-decreasing
numbers satisfying the Cs\"{o}rgo-R\'{e}v\`{e}sz-Stute conditions (S1), (S2)
and (S3), then 
\begin{equation*}
\lim_{N\uparrow +\infty }k_{N}\left( a_{N},\alpha _{N}^{-}\right) =1,a.s.
\end{equation*}

\noindent (II) If 
\begin{equation*}
a_{N}=cN^{-1}\log N,c>0,N\geq 1
\end{equation*}
then 
\begin{equation*}
\lim_{N\uparrow +\infty }k_{N}^{\circ} 
\left( a_{N},\alpha _{N}^{-}\right) =\left( \frac{c}{2}\right) ^{\frac{1}{2}%
}\left( \beta ^{+}-1\right), \text{ } a.s., \text{ }  where \text{ } \beta ^{+}>1
\end{equation*}
and 
\begin{equation*}
\beta ^{+}\left( \log \beta ^{+}-1\right) =c^{-1}-1.
\end{equation*}

\noindent  (III) 
\begin{equation*}
Ifa_{N}=\left( \log N\right) ^{-c},c>0,
\end{equation*}
then 
\begin{equation*}
c^{\frac{1}{2}}\leq \lim_{N\rightarrow +\infty }\inf k_{N}\left(
a_{N},\alpha _{N}^{-}\right) \leq \lim_{N\rightarrow +\infty }\sup
k_{N}\left( a_{N},\bar{\alpha}_{N}\right) \leq \left( 1+c\right) ^{\frac{1}{2%
}},a.s.,
\end{equation*}

\noindent  (IV) If 
\begin{equation*}
a_{N}=c_{N}N^{-1}\log N,c_{N}\rightarrow 0
\end{equation*}

\noindent such that 
\begin{equation*}
\left( c_{N}\log N\right) =Na_{N}\rightarrow +\infty
\end{equation*}

\noindent and 

\begin{equation*}
\left( \log N\right) ^{-1}\left( \log c_{N}^{-1}\right) \log \log
N\rightarrow 0 \text{ } as \text{ } N\uparrow +\infty ,
\end{equation*}

\noindent then

\begin{equation*}
\lim_{N\uparrow +\infty }\sup \frac{N^{\frac{1}{2}}\log \left( \frac{1}{c_{N}%
}\right) }{\log N}\wedge _{N}\left( a_{N},\bar{\alpha}_{N}\right) \leq 2%
\text{, a.s.}
\end{equation*}
\end{theorem}

\noindent We also have

\begin{theorem} \label{tB} If $k=k\left( N\right) \rightarrow +\infty $ such that for some $\delta
>2$ and for some $N_{0},$

\begin{equation*}
0<a_{N}\leq t_{k}\left( \delta \right) =k^{k\left( \delta -2\right) }\exp
\left( \frac{-k^{\delta }}{2}\right) ,\text{ }N\geq N_{0},
\end{equation*}
then Parts (I), (II), (III) and (IV) remain true.
\end{theorem}

\bigskip

\begin{remark} \label{rA} If each $\alpha _{N}$ is the spacings empirical process based on a sample depending
on N, say $\chi\left( N\right) $, and if these samples $\chi\left( N\right) $, $%
N=1,2,...$ are mutually independent (this statistical situation is quite
concievable, for instance when checking homogeneity) the strong limit
results of Theorem \ref{tA} are also valid for $\alpha _{N}$. One might seek other
conditions to get the same extensions. Here, we restrict ourselves to weak
extensions in the following
\end{remark}

\newpage

\begin{corollary} \label{cA} Let k be either fixed or $k\rightarrow +\infty $. Let $\left(
a_{N}\right) _{N\geq 1}$ be a sequence of positive numbers such that $0\leq
a_{N}\leq t_{k}(\delta) $ when $N\geq N_{0},$ for some $N_{0}$ and $\delta >2$. Then:\\

\noindent (I) Under the assumptions of Part I of Theorem \ref{tA}, we have 
\begin{equation*}
\lim_{N\rightarrow +\infty }k_{N}\left( a_{N},\alpha _{N}\right) =1\text{ in
probability}.
\end{equation*}

\noindent (II) Under the assumptions of Part II of Theorem \ref{tA}, we have 
\begin{equation*}
\lim_{N\rightarrow +\infty }k_{N}\left( a_{N},\alpha _{N}\right) =\left( 
\frac{c}{2}\right) ^{\frac{1}{2}}\left( \beta ^{+}-1\right) \text{ in
probability}.
\end{equation*}

\noindent (III) Under the assumptions of Part III of Theorem \ref{tA}, we have 
\begin{equation*}
\lim_{N\rightarrow +\infty }k_{N}\left( a_{N},\alpha _{N}\right) =c^{\frac{1%
}{2}}\text{, in probability}.
\end{equation*}

\noindent (IV) Under the assumptions of Part IV of Theorem \ref{tA}, we have 
\begin{equation*}
\lim_{N\rightarrow +\infty }P\left( \frac{N^{\frac{1}{2}}\log \left( \frac{1%
}{c_{N}}\right) }{\log N}\right) \wedge _{N}\left( a_{N},\alpha _{N}\right)
>\left( 2+\varepsilon \right) =0\text{ for all }\varepsilon >0\text{.}
\end{equation*}
\end{corollary}

\bigskip

\begin{remark} \label{rB}
It appears from Theorems \ref{tA} and \ref{tB} that the oscillation modulus of $\bar{\alpha}%
_{N}$ and that of the uniform empirical process are almost the same. In \cite
{5}, we prove that the exact strong bounds in (I) and (II) reamin for $%
\alpha _{N}$ when $a_{N}$ satisfies further conditions.
\end{remark}

\begin{remark} \label{rC}
One might think that deriving the result of our corollary by using
invariance principles (as given in \cite{1} and \cite{5}) and well-known
results for the Brownian bridge would be easier (at least for some sequences 
$a_{N}$). This is not true at all (see Remark \ref{rD} below).
\end{remark}

\bigskip

\section{Technical lemmas} \label{sec2}

\noindent It will follow from Lemma 1 of section 3 that the increments of $\bar{\alpha}%
_{N}$ behave as the increments of $\gamma _{N}\left( \psi \left( .\right)
\right) $ and those of $\phi \left( .\right) $ where $\gamma _{N}\left(
.\right) $ is the E.P. pertaining to 
\begin{equation*}
U_{1},...,U_{N},\psi \left( s\right) =H_{k}\left( \mu _{n}H_{k}^{-1}\left(
s\right) \right) ,0\leq s\leq 1,\mu _{n}d_{=}\delta _{n},
\end{equation*}
\begin{equation*}
n=1,2,...,\phi \left( s\right) =H_{k}^{\prime }\left( H_{k}^{-1}\left(
s\right) \right) H_{k}^{-1}\left( s\right) ,0\leq s\leq 1,
\end{equation*}

\noindent with $H_{k}^{\prime }\left( x\right) =\frac{dH_{k}\left( x\right) }{dx},$
for all positive x. Then, since $k_{N}\left( .,\gamma N\right) $ is known,
our study is reduced to describing the increments of $\psi \left( .\right) $
and that of $\phi \left( .\right) $, what we do in this paragraph. $\ $

\bigskip

\begin{lemma} \label{A1} Let k be fixed and $a=a_{N}$ be a sequence of positive numbers
satisfying 
\begin{equation}
\left( n^{-1}\log \log n\right) ^{\frac{1}{2}}\log \left( 
\frac{1}{a}\right) \rightarrow \text{ } 0 \text{ } as \text{ } N\rightarrow +\infty \text{ } and \text{ } a\rightarrow 0, \tag{Q1}
\end{equation}

\noindent then as $N\rightarrow ++\infty $, we have the following
properties

\begin{equation}
\sup_{0\leq h\leq a}\sup_{0\leq s\leq 1-h}\left| \psi \left( s+h\right)
-\psi \left( s\right) \right| =a\left( 1+o\left( 1\right) \right) a.s.,
\label{i}
\end{equation}

\noindent uniformly in s, 
\begin{equation}
0\leq s\leq 1-a,\left| \psi \left( s+a\right) -\psi \left( s\right) \right|
=a\left( 1+q\left( a\right) \right) ,  \label{ii}
\end{equation}
where $q\left( a\right) \rightarrow 0$, a.s., as $a\rightarrow 0$.
\end{lemma}

\noindent \textbf{Proof of lemma \ref{A1}}.

\noindent We need several properties of gamma functions. First note that for a fixed k,

\begin{equation}
s=1-H_{k}\left( s\right) =\frac{e^{-x}x^{k-1}}{\left( k-1\right) !}\left\{ 1+%
\frac{k-1}{x}+\frac{k-2}{x^{2}}+...+\frac{\left( k-a\right) !}{x^{k-1}}%
\right\} ,  \label{2.1a}
\end{equation}

\noindent and

\begin{equation*}
x=H_{k}^{-1}\left( 1-s\right) =\log \left( \frac{1}{s}\right) -\log \left(
k-1\right) !+\left( k-1\right) \log x 
\end{equation*}

\begin{equation}
+\log \left( 1+\frac{k-1}{x}+...+\frac{\left( k-1\right) !}{x^{k-1}}\right)   \label{2.1b}
\end{equation}

\noindent from $k-1$ integrations by parts. Next for a fixed k or for $k\rightarrow
+\infty $, we have, as $x\downarrow 0$,

\begin{equation}
s=H_{k}\left( x\right) =\frac{x^{k}}{k!}\left( 1+0\left( x\right) \right) ,
\label{2.2a}
\end{equation}

\noindent and

\begin{equation}
x=H_{k}^{-1}\left( s\right) =\left( k!\right) ^{\frac{1}{k}}\left( 1+O\left( 
\frac{x}{k}\right) \right) ,  \label{2.2b}
\end{equation}

\noindent where for any function $g\left( .\right) $, $g\left( x\right) =O\left(
y\right) $ as $x\downarrow 0$ means that $\lim_{x\downarrow 0}\sup \left| 
\frac{g\left( x\right) }{y}\right| <+\infty$. To see this, use the following inequalities:

\begin{equation*}
0\leq t\leq x\Rightarrow e^{-x}\leq e^{-t}\leq 1,
\end{equation*}

\noindent to obtain that $e^{-x}\frac{x^{k}}{k!}\leq H_{k}\left( x\right) \leq \frac{%
x^{k}}{k!}$ and the results follow. Now, we are able to prove lemma A1.\\

\noindent \underline{Let us continue the proofs Lemma \ref{A1}}. Define

\begin{equation}
\Psi _{h}\left( s\right) =\psi \left( s+h\right) -\psi \left( s\right)
,0\leq s\leq 1-h,0\leq h\leq a,h=1,2,...  \label{2.3}
\end{equation}

\noindent Straighforward computations give

\begin{equation}
\frac{d\Psi _{h}\left( s\right) }{ds}=\mu _{n}^{k-1}\left\{ \exp \left(
\left( \mu _{n}-1\right) H_{k}^{-1}\left( s+h\right) \right) -\exp \left(
\left( \mu _{n}-1\right) H_{k}^{-1}\left( s\right) \right) \right\} .
\label{2.4}
\end{equation}

\noindent Thus, for each elementary event $\omega $ of the probability space, for each
N (that is to say for each n) and for each h, $\Psi _{h}.\left( .\right) $
is non-decreasing of non-increasing according to the sign of $\mu _{n}\left(
\omega \right) -1$. Thus we have

\begin{equation}
\sup_{0\leq s\leq 1-h}\left| \Psi _{h}\left( s\right) \right| =\max \left\{
\Psi _{h}\left( 0\right) ,\left| \Psi _{h}\left( 1-h\right) \right| \right\}
.  \label{2.5}
\end{equation}

\noindent \underline{Computation of $\Psi _{h}\left( 1-h\right)$}. By using (\ref{2.1a}) and (\ref{2.1b}), with $%
h=1-H_{k}\left( x\right) $, we have

\begin{equation*}
\mu _{n}H_{k}^{-1}\left( 1-h\right) =\mu _{n}\log \left( \frac{1}{h}\right)
-\mu _{n}\log \left( k-1\right) !+\mu _{n}\left( k-1\right) \log x
\end{equation*}

\begin{equation*}
+\mu
_{n}\log \left( 1+...+\frac{\left( k-1\right) !}{x^{k-1}}\right)
\end{equation*}

\noindent Now recall that $\Psi _{h}\left( 1-h\right) =1-H_{k}\left( \mu
_{n}H_{k}^{-1}\left( 1-h\right) \right) $ and, using (\ref{2.1a}), get

\begin{equation}
\Psi _{h}\left( 1-h\right) =\mu _{n}^{k-1}\frac{\left( \log \left( \frac{1}{h%
}\right) \right) ^{k-1}h^{\mu _{n}}\left( \left( k-1\right) !\right) ^{\mu
_{n}}}{\left( k-1\right) !}x^{-\left( k-1\right) \mu _{n}}\left(
1+q_{1}\left( h\right) \right) ,  \label{2.6}
\end{equation}

\noindent where there exists $A_{k}$ and $B_{k}$ depending only on k (k being fixed)
such that

\begin{equation}
\left| q_{1}\left( h\right) \right| \leq A_{k}x^{-1}\log
x+B_{k}x^{-1}, \ as \ h\rightarrow 0 \ (i.e. \ as \ x\rightarrow +\infty).  \label{2.7}
\end{equation}

\noindent These constants $A_{k}$ and $B_{k}$ are provided by the approximation

\begin{equation*}
\left| \frac{\log \left( \frac{1}{h}\right) }{x}-1\right| \leq
A_{k}x^{-1}\log x+B_{k}x^{-1},
\end{equation*}

\noindent as $h\rightarrow 0$, $x=H_{k}^{-1}\left( 1-h\right) \rightarrow +\infty$.
And (\ref{2.7}) leads to 

\begin{equation}
\Psi _{h}\left( 1-h\right) =h^{MU_{n}}\left( \log \left( \frac{1}{h}\right)
\right) ^{\left( k-1\right) \left( 1-\mu _{n}\right) }\left( 1+q_{2}\left(
h\right) \right) ^{-\left( k-1\right) \mu _{n}+1}\left( 1+q_{3}\left(
N\right) \right) ,  \label{2.8}
\end{equation}

\noindent where $q_{2}\left( .\right) $ satisfies( \ref{2.7}) with the same
constants A$_{k}$ and B$_{k}$ and $q_{3}\left( N\right) =o\left( 1\right) $,
a.s., independently of $h$, $0\leq h\leq 1$, as $N\rightarrow +\infty $. Since
the functions $x^{-1}\log x$ and $x^{-1}$ are non-increasing as $%
x\rightarrow +\infty $, it follows from (\ref{2.6}), (\ref{2.7}) and (%
\ref{2.8}) that

\begin{equation}
\forall 0\leq h\leq a,\Psi _{h}\left( 1-h\right) =\left( 1+q\left( a\right)
\right) h^{\mu _{n}}\left( \log \left( \frac{1}{h}\right) \right) ^{\left(
k-1\right) \mu _{n}},  \label{2.9}
\end{equation}

\noindent where q$\left( a\right) \rightarrow 0$ and N$\rightarrow +\infty $. By
convention, we shall write $g\left( h\right) =q\left( a\right) $ for $0\leq
h\leq a$, for all h, $0\leq h\leq a$, $g\left( h\right) =o\left( 1\right) 
$ where the ''$o\left( 1\right) $'' depends only on  $a$, as a$\rightarrow
0$.\\

\noindent \underline{Computation of $\Psi _{n}\left( 0\right)$}.

\noindent We have $\Psi _{h}\left( 0\right) =H_{k}\left( \mu _{n}H_{k}^{-1}\right) $. Then by using (\ref{2.2a})-(\ref{2.2b}), we obtain

\begin{equation}
\mu _{n}H_{k}^{-1}\left( h\right) =\mu _{n}\left( k!\right) ^{\frac{1}{k}}h^{%
\frac{1}{k}}\left( 1+q\left( a\right) \right) ,0\leq h\leq a.  \label{2.10}
\end{equation}

\noindent Use again (\ref{2.2a})-(\ref{2.2b}) and get $H_{k}\left( \mu _{n}H_{k}^{-1}\left(
h\right) \right) =h\mu _{n}^{k}\left( 1+q\left( a\right) \right) $, a.s., $%
0\leq h\leq a\rightarrow 0$, since k is fixed and $\mu _{n}\rightarrow 1$,
a.s., as $N\rightarrow +\infty $. Then,

\begin{equation}
\sup_{0\leq h\leq a}\sup_{0\leq s\leq 1-h}\Psi _{h}\left( s\right) =\left(
1+q\left( a\right) \right) \sup_{0\leq h\leq a}\max \left\{ h,h^{\mu
_{n}}\left( \log h^{-1}\right) ^{\left( k-1\right) \left( 1-\mu _{n}\right)
}\right\} ,  \label{2.11}
\end{equation}

\noindent a.s., and $N\rightarrow +\infty $. But,

\begin{equation}
\forall N\geq 1,\frac{d\left\{ h^{\mu _{n}}\left( \log h^{-1}\right)
^{\left( k-1\right) \left( 1-\mu _{n}\right) }\right\} }{dh}=h^{\mu
_{n}}\left( \log h^{-1}\right) ^{\left( k-1\right) \left( 1-\mu _{n}\right)
} 
\end{equation}

\begin{equation}
\times \left\{ \mu _{n}-\frac{\left( k-1\right) \left( 1-\mu _{n}\right) }{\log
h^{-1}}\right\} .  \label{2.12}
\end{equation}

\noindent Thus $h^{\mu _{n}}\left( \log h^{-1}\right) ^{\left( k-1\right) \left( \mu
_{n}-1\right) }$ is non-decreasing when $n$ sufficiently large since $k\left(
1-\mu _{n}\right) \rightarrow 0$, a.s., as $N\rightarrow +\infty $ by the
strong law of large numbers ($k$ being fixed).Then,

\begin{equation}
0\leq h\leq a\Rightarrow h^{\mu _{n}}\left( \log h^{-1}\right) ^{\left(
k-1\right) \left( 1-\mu _{n}\right) }\leq a^{\mu _{n}}\left( \log
a^{-1}\right) ^{\left( k-1\right) \left( 1-\mu _{n}\right) },  \label{2.13}
\end{equation}

\noindent a.s., as  $N\rightarrow +\infty$. Furthermore,

\begin{equation}
\left( \log a^{-1}\right) ^{\left( k-1\right) \left( 1-\mu _{n}\right)
}=\exp \left( \left( k-1\right) \left( 1-\mu _{n}\right) \log \log
a^{-1}\right) =1+o\left( 1\right) ,  \label{2.14}
\end{equation}

\noindent a.s., whenever $\left( 1-\mu _{n}\right) \log \log a^{-1}\rightarrow 0$,
a.s. But this is implied by (Q1). Indeed, we have by the law of the
iterated logarithm (the loglog law) that

\begin{equation}
\lim_{N\rightarrow +\infty }\sup \left( 2n^{-1}\log \log n\right) ^{\frac{1}{%
2}}\left| \mu _{n}-1\right| \leq 1, \ a.s.  \label{2.15}
\end{equation}

\noindent This together with (Q1) imply that $\left( 1-\mu _{n}\right) \log \log
a^{-1}\rightarrow 0$, a.s., as $N\rightarrow +\infty $. In fact, the loglog
law holds for $\delta _{n}$, that is

\begin{equation}
\lim_{N\rightarrow +\infty }\sup \left( 2n^{-1}\log \log n\right) ^{\frac{1}{%
2}}\left| \delta _{n}-1\right| \leq 1,a.s.  \label{2.16}
\end{equation}

\noindent But (\ref{2.16}) may be obtained from (see \cite{5}, Appendix)

\begin{equation*}
\sum_{p\geq 0}P\left( \bigcup_{n_{p}}^{n_{p+1}-1}\left\{ \left( 2n^{-1}\log
\log n\right) \left| \delta _{n}-1\right| \geq 1+\frac{\varepsilon }{2}%
\right\} \right) <+\infty ,
\end{equation*}

\noindent where $\left( n_{p}\right) $ is an increasing and unbounded sequence of
positive integers and $\varepsilon >0$ is arbitrary. This and the equality
in distribution of $\delta _{n}$ and $\mu _{n}$ for each N imply (\ref{2.15}). The same $loglog-law$ shows that (Q1) implies that

\begin{equation}
a^{1-\mu _{n}}=\exp \left( \left( 1-\mu _{n}\right) \log a^{-1}\right)
=1+o\left( 1\right) asN\rightarrow +\infty .  \label{2.17}
\end{equation}

\noindent We finally get from (\ref{2.9}), (\ref{2.13}), (\ref{2.14}) and (\ref{2.17}) that

\begin{equation}
\sup_{0\leq h\leq a}\sup_{0\leq s\leq 1-h}\left| \Psi _{h}\left( s\right)
\right| =a\left( 1+o\left( 1\right) \right) a.s.,adN\rightarrow +\infty ,
\label{2.18}
\end{equation}

\noindent which proves Part (\ref{i}) of Lemma \ref{A1}. To prove part (\ref{ii}), it
suffices to remark that we may have through (\ref{2.4}) that 

\begin{equation*}
\min \left( \phi _{a}\left( 0\right) ,\phi _{a}\left( 1-a\right) \right)
\leq \phi _{a}\left( s\right) \leq \max \left( \phi _{a}\left( 0\right)
,\phi _{a}\left( 1-a\right) \right) ,0\leq s\leq -a,
\end{equation*}

\noindent and the part in question follows since the first part implies that $\phi
_{a}\left( 0\right) =a\left( 1+o\left( 1\right) \right) $, a.s. and $\phi
_{a}\left( 1-a\right) =a\left( 1+o\left( 1\right) \right) $, a.s., as $%
N\rightarrow +\infty $.

\bigskip

\begin{lemma} \label{A2} Let k be fixed, then we have as $a\rightarrow 0$, $N\rightarrow +\infty
, $%
\begin{equation*}
\sup_{0\leq h\leq a}\sup_{0\leq s\leq 1-h}\left| \phi \left( s\right) -\phi
\left( s+h\right) \right| =\left( a\log a^{-1}\right) \left( 1+o\left(
1\right) \right) .
\end{equation*}
\end{lemma}

\bigskip

\noindent \textbf{Proof of Lemma \ref{A2}}.

\noindent Consider $\Phi _{h}\left( s\right) =\phi \left( s+h\right) -\phi \left(
s\right) $, $0\leq s\leq 1-h$. Direct considerations yield that 
\begin{equation*}
\frac{d\Phi _{h}\left( s\right) }{ds}=H_{k}^{-1}\left( s\right)
-H_{k}^{-1}\left( s+h\right) ,0\leq s\leq 1-h.
\end{equation*}

\noindent Then for each h, $\Phi _{h}\left( .\right) $ is non-increasing and thus,

\begin{equation*}
\sup_{0\leq s\leq 1-h}\left| \Phi _{h}\left( s\right) \right| =\max \left\{
\left| \Phi _{h}\left( 0\right) \right| ,\text{ }\left| \Phi _{h}\left(
1-h\right) \right| \right\} .
\end{equation*}

\noindent But, by (\ref{2.2a})-(\ref{2.2b}), 
\begin{equation*}
\Phi _{h}\left( 0\right) =H_{k}^{\prime }\left( H_{k}^{-1}\left( h\right)
\right) H_{k}^{-1}\left( h\right) =kh\left( 1+q\left( a\right) \right) ,
\end{equation*}

\noindent $0\leq h\leq a\rightarrow 0$. Here we omit the details concerning the
uniform approximations which provide $q\left( .\right) $. These details are
very similar to those of the computation of $\Phi _{h}\left( 0\right) $.
By the considerations that were previously used for getting (\ref{2.6}) from (\ref
{1.1}), we have

\begin{equation*}
\Phi _{h}\left( 1-h\right) =H_{k}^{\prime }\left( H_{k}^{-1}\left(
1-h\right) \right) H_{k}^{-1}\left( 1-h\right) =h\log h^{-1}\left( 1+q\left(
a\right) \right) ,0\leq h\leq a\rightarrow 0.
\end{equation*}

\noindent Notice that $H_{k}^{\prime }\left( H_{k}^{-1}\left( 1-h\right) \right) $
yields something like (\ref{2.6}) while $H_{k}^{-1}\left( 1-h\right) $
yields $\left( \log h^{-1}\right) \left( 1+q\left( a\right) \right) $, $%
0\leq h\leq a\rightarrow 0$. We obtain 
\begin{equation*}
\sup_{0\leq h\leq a}\sup_{0\leq s\leq 1-h}\left| \Phi _{h}\left( s\right)
\right| =1+q\left( a\right) \sup_{0\leq h\leq a}\max \left( kh\text{, }h\log
h^{-1}\right) =\left( 1+q\left( a\right) \left( a\log a^{-1}\right) \right) ,
\end{equation*}

\noindent $a\rightarrow 0$, since $k$ is fixed here. Hence Lemma \ref{A2} is proved.

\bigskip

\noindent Now, we concentrate on the case where $k\rightarrow +\infty $, First, we give
the following

\bigskip

\begin{proposition} \label{p1} Let 
\begin{equation*}
0\leq s\leq t_{k}\left( \delta \right) =k^{k\left( \delta -2\right) }\exp
\left( -\frac{1}{2}k^{\delta }\right) ,
\end{equation*}

\noindent Then, as $k\rightarrow +\infty ,$ we have 
\begin{equation}
x=H_{k}^{-1}\left( 1-s\right) =\left( \log s^{-1}\right) \left(
1+q_{4}\left( s\right) \right) ,  \label{2.19}
\end{equation}

\noindent where there exist A and $k_{0}$ such that $\left| q_{4}\left( s\right)
\right| \leq A\frac{\log k}{k^{\delta -1}}$ for all $0\leq s\leq t_{k}\left(
\delta \right) ,$ $k\geq k_{0}$.
\end{proposition}

\bigskip

\begin{proof} Integrating by parts, we get 
\begin{equation*}
\forall x\geq 0,\frac{x^{k-1}e^{-x}}{\left( k-1\right) !}\leq 1-H_{k}\left(
x\right) \leq \frac{x^{k-1}e^{-x}}{\left( k-1\right) !}+\frac{k}{x}\left(
1-H_{k}\left( x\right) \right)
\end{equation*}

\noindent Then, 
\begin{equation}
\frac{x^{k-1}e^{-x}}{\left( k-1\right) !}\leq 1-H_{k}\left( x\right) \leq
\left\{ 1-\frac{k}{x}\right\} ^{-1}\frac{x^{k-1}e^{-x}}{\left( k-1\right) !}%
,\forall x\geq 0.  \label{2.20}
\end{equation}

\noindent We are able to see that the expansion of $H_{k}\left( x\right) $ is then
possible if $k/x\rightarrow 0$. Now, let $0\leq s\leq s_{k}=1-H_{k}\left(
k^{\delta }\right) $. Apply (\ref{2.20}) and get 

\begin{equation*}
0\leq s\leq s_{k}\Rightarrow s=1-H_{k}\left( x\right) =\frac{x^{k-1}e^{-x}}{%
\left( k-1\right) !}\left\{ 1+q_{5}\left( x\right) \right\} ,
\end{equation*}

\noindent with $\left| q_{5}\left( x\right) \right| \leq \left( 1-k^{-\left( \delta
-1\right) }\right) k^{-\left( \delta -1\right) }$ for all $0\leq s\leq s_{k}$%
. But 

\begin{equation*}
s_{k}=\frac{k^{\delta \left( k-1\right) }e^{-k}}{\left( k-1\right) !}\left(
1+0\left( k^{-\left( \delta -1\right) }\right) \right) .
\end{equation*}

\noindent Then by Sterling's formula and some sthraighforward calculations, it is
possible to find a $k_{1}$ such that $t_{k}\left( \delta \right) =k^{k\left(
\delta -2\right) }\exp \left( -\frac{1}{2}k^{\delta }\right) \leq s_{k}$ for
all $k\leq k_{1}$. Then for $0\leq s\leq s_{k},k\geq k_{1},$%
\begin{equation}
x=H_{k}^{-1}\left( 1-s\right) =\left( \log s^{-1}\right) -\log \left(
k-1\right) !+\left( k-1\right) \log x+o\left( q_{5}\left( s\right) \right) .
\label{2.21}
\end{equation}

\noindent Now since 
\begin{equation*}
0\leq s\leq s_{k}\Rightarrow \left| \frac{\left( k-1\right) \log x}{x}%
\right| \leq \frac{\log k}{k^{\delta -1}}=0\left( k^{-\left( \delta
-1\right) }\log k\right) ,
\end{equation*}

\begin{equation*}
0\leq s\leq s_{k}\Rightarrow \left| \frac{\log \left( k-1\right) !}{x}%
\right| \leq \frac{\log k!}{k^{\delta }}=0\left( k^{-\left( \delta -1\right)
}\log k\right) ,
\end{equation*}

\noindent by Sterling's formula. Thus, these two facts and (\ref{2.21}) together imply
that 
\begin{equation*}
\forall s\leq s_{k},\log s^{-1}=x\left( 1+0\left( k^{-\left( \delta
-1\right) }\log k\right) =H_{k}^{-1}\left( 1-s\right) (1+0\left( k^{-\delta
+1}\log k\right) \right) ,
\end{equation*}

\noindent which was to be proved. We finally give two lemmas which correspond to
Lemmas \ref{A1} and \ref{A2} in the case of infinite steps $k$.
\end{proof}

\bigskip

\begin{lemma} \label{A3} Let k satisfy, as $ N\rightarrow +\infty$, 

\begin{equation}
\left( K\right) kN^{-1}\left( \log \log n\right)\rightarrow 0 \tag{K}
\end{equation}

\noindent and 
\begin{equation}
k=k\left( N\right) \rightarrow +\infty \tag{Q2}
\end{equation}

\noindent Then the following assertions hold.

\begin{equation}
\sup_{0\leq h\leq a}\sup_{0\leq s\leq 1-h}\left| \psi \left( s+h\right)
-\psi \left( s\right) \right| =a\left( 1+o\left( 1\right) \right) \text{,
a.s., as }N\rightarrow +\infty .
\end{equation}

\begin{equation}
\left| \psi \left( s+h\right) -\psi \left( s\right) \right| =a\left(
1+q\left( a\right) \right) ,\text{for }0\leq s\leq 1-a\text{ , with }q\left(
a\right) \rightarrow 0\text{, a.s., as }N\rightarrow +\infty .
\end{equation}
\end{lemma}

\bigskip

\noindent \textbf{Proof of of Lemma \ref{A3}}. As in Lemma \ref{A1}, we have 
\begin{equation}
\sup_{0\leq s\leq 1-h}\left| \psi _{h}\left( s\right) \right| =\max \left\{
\left| \psi _{h}\left( 0\right) \right| ,\left| \psi _{h}\left( 1-h\right)
\right| \right\} .  \label{2.22}
\end{equation}

\noindent First we treat $\psi _{h}\left( 0\right) =H_{k}\left( \mu
_{n}H_{k}^{-1}\left( h\right) \right) .$ Equations (\ref{2.2a})-(\ref{2.2b}) yield 
\begin{equation*}
\mu _{n}H_{k}^{-1}\left( h\right) =\left( k!\right) ^{\frac{1}{k}}\left(
1+o\left( H_{k}^{-1}\frac{\left( a\right) }{k}\right) \right) \mu _{n},0\leq
h\leq a.
\end{equation*}

\noindent Now we note that $0\leq s\leq a$ implies that $0\leq H_{k}^{-1}\left(
h\right) \leq C_{1}a^{\frac{1}{k}}\left( k!\right) ^{\frac{1}{k}}$ for small
values of a, $C_{1}$ being a constant. Sterling's formula then implies for
large values of k, 
\begin{equation*}
0\leq s\leq t_{k},0\leq H_{k}^{-1}\left( h\right) \leq Const.\text{ }%
k^{\delta -1}\exp \left( -\frac{1}{2}k^{\delta -1}\right) .
\end{equation*}
Then $H_{k}^{-1}\left( h\right) \rightarrow 0$ and we are able to use (\ref
{2.2a})-\ref{2.2b} to get 
\begin{equation*}
\forall 0\leq h\leq a\Rightarrow H_{k}\left( \mu _{n}H_{k}^{-1}\left(
h\right) \right) =\mu _{n}^{k}h\left( 1+o(H_{k}^{-1}\left( a\right) \right) .
\end{equation*}

\noindent The $loglog-law$ implies that 
\begin{equation*}
k\left( 1-\mu _{n}\right) =0\left( k\left( 2n^{-1}\log \log n\right)
^{2}\right) ,a.s.
\end{equation*}
Thus, whenever $\left( K\right) $ is satisfied, one has 
\begin{equation*}
\mu _{n}^{k}=\exp (-k\left( 1-\mu _{n}\right) \left( 1+o\left( 1\right)
\right) \rightarrow 1,\text{a.s.}
\end{equation*}

\noindent Hence 
\begin{equation}
\forall 0\leq s\leq a\leq t_{k}\left( \delta \right) ,\psi _{h}\left(
0\right) =h\left( 1+q\left( a\right) \right) \leq a\left( 1+q\left( a\right)
\right) =\psi _{a}\left( 0\right) .  \label{2.23}
\end{equation}

\noindent We now treat $\psi _{h}\left( 1-h\right) .$ By the proposition, we get 
\begin{equation}
\forall 0\leq h\leq a\leq t_{k}\left( \delta \right) ,\text{ }X=\mu
_{n}H_{k}^{-1}\left( 1-h\right) =\mu _{n}\left( \log h^{-1}\right) \left(
1+0\left( \frac{\log k}{k^{\delta -1}}\right) \right) ,a.s.  \label{2.24}
\end{equation}

\noindent Since $\frac{X}{k}=0\left( k^{-\left( \delta -1\right) }\right) ,a.s.,$ one
has 
\begin{equation}
1-H_{k}\left( X\right) =\mu _{n}^{k-1}\frac{\left( \log h^{-1}\right) ^{k-1}%
}{\left( k-1\right) !}\left( 1+0\left( \frac{\log k}{k^{\delta -2}}\right)
\right) h^{\mu _{n}}x^{-\left( k-1\right) \mu _{n}}\left( \left( k-1\right)
!\right) ^{\mu _{n}}\left( 1+q_{4}\left( h\right) \right) ,a.s.,
\label{2.25}
\end{equation}

\noindent as $N\rightarrow +\infty $. Replace x by $\log h^{-1}$ in (\ref{2.25}). On
account of (\ref{2.24}) and of the fact that $\left( 1+0\left( k^{1-\delta
}\log k\right) ^{k-1}\right) =\left( 1+0\left( k^{2-\delta }\log k\right)
\right) $, we get 
\begin{equation*}
\psi _{h}\left( 1-h\right) =\left( \left( k-1\right) !\right) ^{1-\mu
_{n}}\left( \log h^{-1}\right) ^{\left( k-1\right) \left( 1-\mu _{n}\right)
}h^{\mu _{n}}\left( 1+q\left( a\right) \right) .
\end{equation*}

\noindent Finally, by taking $(K)$ and $(Q2)$ into account, we find
ourselves in the same situation as in the proof of Lemma \ref{A1} (see Statement (%
\ref{2.8})). But in order to have the same conclusion, i.e., 
\begin{equation}
\sup_{0\leq h\leq a}\psi _{h}\left( 1-h\right) =a\left( 1+q\left( a\right)
\right) ,a.s.,\text{ as }N\rightarrow +\infty ,  \label{2.27}
\end{equation}

\noindent we have to check that 
\begin{equation*}
\left( \left( k-1\right) !\right) ^{1-\mu _{n}}=\exp \left( \left( 1-\mu
_{n}\right) \log \left( k-1\right) !\right) =:\rho _{n}\rightarrow
1,a.s.,asN\rightarrow +\infty .
\end{equation*}

\noindent But the $loglog-law$ and Sterling's formula together show that 
\begin{equation*}
\rho _{n}=\exp \left( 0\left( \left( \frac{k^{\frac{1}{4}}\left( \log \log
n\right) ^{\frac{1}{2}}}{N^{\frac{1}{2}}}\log a^{-1}\right) \left( \frac{%
\log k}{\log a^{-1}}\right) \right) \right) .
\end{equation*}

\noindent Obviously the condition $0<a\leq t_{k}\left( \delta \right) $ implies that $%
(log k)/(\log a^{-1})\rightarrow 0$ as $N\rightarrow +\infty $, and as $%
k\rightarrow +\infty $. This fact combined with (Q2) clearly shows that $%
\rho _{n}\rightarrow 1$ as $N\rightarrow +\infty $. Now, by putting together (\ref
{2.22}), (\ref{2.23}) and (\ref{2.27}), we get 
\begin{equation}
\left( 0<a\leq t_{k}\left( \delta \right) ,\delta >2\right) \Rightarrow
\sup_{0\leq 0\leq a}\sup_{0\leq s\leq 1-h}\left| \psi _{h}\left( s\right)
\right| =a\left( 1+q\left( a\right) \right) ,a.s.,\text{ as }N\rightarrow
+\infty .  \label{2.28}
\end{equation}

\bigskip

\begin{lemma} \label{A4} Let $0<a\leq t_{k}\left( \delta \right) ,\delta >2$. Then as $k\rightarrow +\infty$, we have 
\begin{equation*}
\sup_{0\leq h\leq a}\sup_{0\leq s\leq 1-h}\left| \phi \left( s+h\right)
-\phi \left( s\right) \right| =\left( a\log a^{-1}\right) \left( 1+q\left(
a\right) \right) ,q\left( a\right) \rightarrow 0\text{ as }N\rightarrow
+\infty ,a\rightarrow 0.
\end{equation*}
\end{lemma}

\bigskip

\noindent \textbf{Proof of of Lemma \ref{A4}}. If we proceed as in Lemma A2 and as in Lemma A3, we get 
\begin{equation*}
\sup_{0\leq h\leq a}\sup_{0\leq s\leq 1-h}\left| \Phi _{h}\left( s\right)
\right| =\max \left( ka,\left( a\log a^{-1}\right) \right) \left( 1+q\left(
a\right) \right) ,\text{ as }N\rightarrow +\infty ,\text{ }a\rightarrow 0
\end{equation*}

\noindent From there, the conclusion is obtained by noticing that the condition $%
0<a\leq t_{k}\left( \delta \right) $ implies that $\frac{\left( \log
a^{-1}\right) }{k}\rightarrow +\infty $ as $k\rightarrow +\infty .$

\bigskip

\section{Proofs of the results}.

\noindent Throughout, we shall use the following representation which follows from 
\cite{5} (see e.g. the study of $R_{N1}\left( x\right) $).

\begin{lemma} \label{lem1} Let k be fixed of $k\rightarrow +\infty $ as $N\rightarrow +\infty $%
, then 
\begin{equation*}
\left\{ \alpha _{N}\left( s\right) ,0\leq s\leq 1\right\} d_{=}\left\{
\gamma _{N}\left( \psi \left( s\right) \right) +N^{\frac{1}{2}}\left\{
H_{k}\left( \mu _{n}H_{k}^{-1}\left( s\right) \right) -s\right\} ,0\leq
s\leq 1\right\} 
\end{equation*}

\begin{equation*}
 =:\left\{ \bar{\alpha}_{N}\left( s\right) ,0\leq s\leq
1,a.s.\right\} 
\end{equation*}
\end{lemma}

\noindent Lemma \ref{lem1} will be systematically used. Then, if $a_{N}$ satisfies

\begin{equation}
\lim_{N\rightarrow +\infty }\frac{\left( \log \log n\right) ^{2}}{Na_{N}\log
a_{N}^{-1}}=0,  \label{Q3}
\end{equation}

\noindent we will be able to focus our attention on $\gamma _{N}\left( \psi \left(
s\right) \right) +N^{\frac{1}{2}}\left( \mu _{n}-1\right) \phi \left(
s\right) $ in the following way
\begin{equation}
\frac{\bar{\alpha}_{N}\left( s\right) }{b_{N}}=\frac{\gamma _{N}\left( \psi
\left( s\right) \right) }{b_{N}}+\frac{N^{\frac{1}{2}}\left( \mu
_{n}-1\right) \phi \left( s\right) }{b_{N}}+b_{N}^{-1}0\left( N^{\frac{1}{2}%
}\log \log n\right) ,a.s.  \label{3.1}
\end{equation}
\begin{equation*}
=:A_{N1}\left( s\right) +A_{N2}\left( s\right) +A_{N3}\left( s\right) .
\end{equation*}
with $b_{N}=\left( 2a_{N}\log \log a_{N}^{-1}\right) ^{\frac{1}{2}}=b\left(
a_{N}\right) .$ It follows that if $(Q3)$ holds we have $A_{N3}\left( s\right)
=o\left( 1\right) ,a.s.,$ uniformly with respect to $s$, $0\leq s\leq 1$.\\

\bigskip 

\noindent \textbf{Proof of Part I of Theorem \ref{tA}}. By (\ref{2.28}), we have
\begin{equation}
k_{N}\left( a_{N},\bar{\alpha}_{N}\right) \leq k_{N}\left( a_{N},\text{ }%
A_{N1}\right) +k_{N}\left( a_{N},\text{ }A_{N2}\right) +k_{N}\left( a_{N},%
\text{ }A_{N3}\right) ,  \label{3.2}
\end{equation}

\noindent and by Lemma \ref{A3}, we have for a fixed $k$, 
\begin{equation*}
k_{N}\left( a_{N},\text{ }A_{N2}\right) \leq N^{\frac{1}{2}}\left| 1-\mu
_{n}\right| b\left( a_{N}\right) \left( 1+o\left( 1\right) \right) ,a.s.,
\end{equation*}

\noindent as $N\uparrow +\infty $. Thus the $loglog-law$ implies that
\begin{equation}
\lim_{N\rightarrow +\infty }k_{N}\left( a_{N},\text{ }A_{N2}\right) =o\left(
1\right) ,a.s.,  \label{3.3}
\end{equation}

\noindent whenever
\begin{equation}
k^{-\frac{1}{2}}\left( 2\log \log n\right) ^{\frac{1}{2}}b\left(
a_{N}\right) \rightarrow 0\text{ as }N\rightarrow +\infty   \tag{Q4}
\end{equation}

\noindent is satisfied. On the other hand, Lemma \ref{A1} and Theorem $0.2$ of Stute \cite{10}
together yield that
\begin{equation}
k_{N}\left( a_{N},A_{N1}\right) \leq k_{N}\left( a_{N},\text{ }\gamma
_{N}\right) \left( 1+o\left( 1\right) \right) =1+o\left( 1\right) ,a.s.,%
\text{ as }N\rightarrow +\infty   \label{3.4}
\end{equation}

\noindent Then if $(Q1)$, $(Q3)$, $(Q4)$, $(S1)$ and $(S3)$ are satisfied, we
get
\begin{equation*}
\lim_{N\rightarrow +\infty }k_{N}\left( a_{N},\text{ }\bar{\alpha}%
_{N}\right) \leq 1,a.s.
\end{equation*}

\noindent Now let
\begin{equation*}
\theta _{N}\left( a_{N},\text{ }R_{N}\right) =\sup_{0\leq s\leq
1-a_{N}}\left\{ R_{N}\left( s+a_{N}\right) -R_{N}\left( s\right) \right\} .
\end{equation*}

\noindent By Lemma \ref{A2}, we have for large $N$ that

\begin{equation}
b_{N}^{-1}\sup_{0\leq s\leq 1-a_{N}}\left| A_{N2}\left( s+a_{N}\right)
-A_{N2}\left( s\right) \right| \leq N^{\frac{1}{2}}\left| 1-\mu _{n}\right|
b\left( a_{N}\right) \left( 1+o\left( 1\right) \right) ,a.s.  \label{3.5}
\end{equation}

\noindent Thus if $(Q3)$ and $(Q4)$ are satisfied, we get
\begin{equation}
\theta _{N}\left( a_{N},\text{ }\bar{\alpha}_{N}\right) \leq \theta
_{N}\left( a_{N},\text{ }A_{N1}\right) +o\left( 1\right) ,a.s.,\text{ as }%
N\rightarrow +\infty .  \label{3.6}
\end{equation}

\noindent Furthermore it may be derived from Theorem $0.2$ of Stute \cite{10} that $(S1)$,
$(S2)$ and $(S3)$ yield
\begin{equation}
b_{N}^{-1}\sup_{0\leq s\leq 1-a_{N}}\left| \gamma _{N}\left( \psi \left(
s\right) +o\left( a_{N}\right) \right) -\gamma _{N}\left( \psi \left(
s\right) +a_{N}\right) \right| =o\left( 1\right) ,a.s.,\text{ as }%
N\rightarrow +\infty   \label{3.7}
\end{equation}

\noindent It follows from (\ref{3.6}) and (\ref{3.7}) that $(S1-2-3)$ and $(Q1-3-4)$
together imply
\begin{equation*}
b_{N}^{-1}\theta _{N}\left( a_{N},\text{ }\bar{\alpha}_{N}\right) \geq
b_{N}^{-1}\left\{ \sup_{0\leq s\leq 1-a_{N}}\gamma _{N}\left( \psi \left(
s\right) +a_{N}\right) -\gamma _{N}\left( \psi \left( s\right) \right)
\right\} +o\left( 1\right) \ a.s,\end{equation*}

\noindent  as $N\rightarrow +\infty$. Since $\psi :\left( 0,\text{ }1-a_{N}\right) \rightarrow \left( 0,\text{ }%
\psi \left( 1-a_{N}\right) \right) $, is a bijection and since $\psi \left(
1-a_{N}\right) =1-a_{N}\left( 1+o\left( 1\right) \right) ,a.s.$, we may use
Lemma \ref{A1} (formulas (\ref{2.4}) and (\ref{2.5}) when $(Q1)$ holds to
find for any $\varepsilon >0$, for any elementary event $\omega $, an $%
N_{o}\left( \omega \right) $ such that
\begin{equation*}
N>N_{o}\Rightarrow b_{N}^{-1}\theta _{N}\left( a_{N},\text{ }\bar{\alpha}%
_{N}\right) \geq \sup_{0\leq s\leq 1-a_{N}\left( 1+\varepsilon \right)
}\left( \gamma _{N}\left( s+a_{N}\right) -\gamma _{N}\left( s\right) \right)
+o\left( 1\right) ,a.s.
\end{equation*}

\noindent Once again, we use the Theorem 02 of \cite{10} to see that, under (S1-2-3),
we have

\begin{equation*}
\lim_{N\rightarrow +\infty }\sup_{0\leq s\leq 1-a_{N}\left( 1+\varepsilon
\right) }\left\{ \frac{\left| \gamma _{N}\left( s+a_{N}+\varepsilon
a_{N}\right) -\gamma _{N}\left( s+a_{N}\right) \right| }{b_{N}}\right\} \leq
\varepsilon ^{\frac{1}{2}},a.s.
\end{equation*}

\noindent Thus, under (Q1-3-4) and for large values of N, we get
\begin{equation}
b_{N}^{-1}\theta _{N}\left( a_{N},\text{ }\bar{\alpha}_{N}\right) \geq
b_{N}^{-1}\theta _{N}\left( a_{N},\text{ }\gamma _{N}\right) -\left( 1+\frac{%
1}{2}\right) \varepsilon ^{\frac{1}{2}},a.s.,  \label{3.8}
\end{equation}

\noindent Hence Lemma \ref{2.9} in \cite{10} and (\ref{3.8}) together yield
\begin{equation}
\forall \varepsilon >0,\lim_{N\rightarrow +\infty }\inf \theta _{N}\left(
a_{N},\text{ }\bar{\alpha}_{N}\right) \geq \left( 1-\varepsilon \right) ^{%
\frac{1}{2}}-\left( 1+\frac{1}{2}\right) ,a.s.  \label{3.9}
\end{equation}

\noindent Finally (\ref{3.4}) and (\ref{3.9}) together ensure that
\begin{equation*}
\lim_{N\rightarrow +\infty }k_{N}\left( a_{N},\text{ }\bar{\alpha}%
_{N}\right) =1,a.s.,
\end{equation*}

\noindent whenever $(Q1-3-4)$ and $(S1-2-3)$ hold. But since $\log n\sim \log N$ (k being
fixed), one has 

\begin{equation}
\left( \frac{\log \log n}{n}\right) ^{\frac{1}{2}}\log a_{N}^{-1}\sim k^{%
\frac{1}{2}}\left( \frac{\log \log N}{\log a_{N}^{-1}}\right) ^{\frac{1}{2}}%
\frac{\left( a_{N}^{\frac{1}{2}}\log a_{N}^{-1}\right) ^{\frac{1}{2}}}{%
\left( N^{\frac{1}{2}}a_{N}^{\frac{1}{2}}\right) },
\end{equation}

\begin{equation}
\left( \frac{\log \log N}{Na_{N}\log a_{N}^{-1}}\right) ^{2}\log
a_{N}^{-1}\leq \frac{\left( \log \log N\right) ^{2}}{\left( \log
a_{N}^{-1}\right) ^{2}},
\end{equation}

\begin{equation}
\left( 2\log \log n\right) \left( a_{N}\log a_{N}^{-1}\right) \sim \left( 
\frac{\log \log N}{\log a_{N}^{-1}}\right) \left( a_{N}^{\frac{1}{2}}\log
a_{N}^{-1}\right) ^{\frac{3}{2}},  \label{iii}
\end{equation}

\noindent for large N. (\ref{i}), (\ref{ii}) and (\ref{iii}) show that (S1) and (S2)
imply (Q1-2-3) and this completes the proof of part I of Theorem \ref{tA}.\\

\bigskip 
\noindent \textbf{Proof of Part II of Theorem \ref{tA}}.\\

\noindent The proof is the same as that of the first part. We only notice that if $%
a_{N}=cN^{-1}\log N,$ $c>0$, $(Q1-3-4)$ are satisfied for a fixed k. To get
Part II of Theorem \ref{tA}, we use Theorem 1 (Part I) of \cite{7} for the
inequality $"\leq "$ and the Erd\"{o}s-Renyi law for the increments of the
uniform empirical process due to Komlos \textit{et al.}, and \cite{4} for the inequality $"\geq "$. Similarly to the first case, we get an analogue to (\ref{3.8}). That is, for any $\varepsilon >0$, for any
elementary event $\omega $, we can find an $N_{1}\left( \omega \right) $
such that

\begin{equation*}
N>N_{1}\Rightarrow k_{N}\left( a_{N},\text{ }\bar{\alpha}_{N}\right) \geq
\sup_{0\leq s\leq 1-a\left( 1+\varepsilon \right) }\left| \frac{\gamma
_{N}\left( s+a\left( 1+\varepsilon \right) \right) -\gamma _{N}\left(
s\right) }{b_{N}}\right| 
\end{equation*}

\begin{equation}
-2\varepsilon ^{\frac{1}{2}}h\left( c\varepsilon
\right) +\varepsilon ^{\frac{1}{2}},   \label{3.10}
\end{equation}

\noindent where for any s, $h\left( s\right) =\left( \frac{s}{2}\right) ^{\frac{1}{2}%
}\left( \beta ^{+}\left( s\right) -1\right) $ and $\beta ^{+}\left( s\right) 
$ is the unique solution of the equation $x\left( \log x-1\right) +1=s^{-1}$
such that $\beta ^{+}\left( s\right) \geq 1.$ Now, since for any $f\left(
.\right) $, $g\left( .\right) $, K, 

\begin{equation*}
\sup_{seK}\max \left( f\left( s\right) ,\text{ }g\left( s\right) \right)
=\max \left( \sup_{s \in K}f\left( s\right) ,\text{ }\sup_{s \in K}g\left( s\right)
\right) \text{ and }\left| x\right| =\max \left( x,-x\right) 
\end{equation*}

\noindent and since (see e.g. the third formula that follows Statement 11 in \cite{7})
\begin{equation*}
\forall \varepsilon >0,\lim_{N\rightarrow +\infty }\inf \frac{\left| \gamma
_{N}\left( s+a\left( 1+\varepsilon \right) \right) -\gamma _{N}\left(
s\right) \right| }{b_{N}}=\left( 1+\varepsilon \right) ^{\frac{1}{2}}h\left(
\left( 1+\varepsilon \right) c\right) ,a.s.,
\end{equation*}

\noindent then (\ref{3.10}) implies that

\begin{equation*}
\forall \varepsilon >0,\lim_{N\rightarrow +\infty }\inf k_{N}\left( a_{N},%
\text{ }\bar{\alpha}_{N}\right) \geq \left( 1+\varepsilon \right) ^{\frac{1}{%
2}}h\left( \left( 1+\varepsilon \right) c\right) -2^{\frac{1}{2}}h\left(
c\varepsilon \right) -\varepsilon ^{\frac{1}{2}},a.s.
\end{equation*}

\bigskip

\noindent Thus it suffices to prove that : (\ref{i}) for each fixed c, $h\left( \left(
1+\varepsilon \right) c\right) \rightarrow h\left( c\right) $ as $%
\varepsilon \rightarrow 0$, and : (\ref{ii}) for each fixed c, $\varepsilon
^{\frac{1}{2}}h\left( c\varepsilon \right) \rightarrow 0$ as $\varepsilon
\rightarrow 0$. But these two points may be directly obtained by simple
considerations.\\

\noindent \textit{Proof of part III of Theorem \ref{tA}}.\\
\\
The proof is very similar to that of Part I of Theorem \ref{tA}. If suffices to
remark that part III of Theorem 1 in \cite{7} holds in the general case
where $a_{N}=\alpha \left( \log N\right) ^{-c},c>0,\alpha >0$.\\

\textit{Proof of Part IV of Theorem \ref{tA}}.\\

\noindent Here $a_{N}=c_{N}N^{-1}\log N,c_{N}\rightarrow 0$ as $N\rightarrow +\infty$. Let $d_{N}=N^{\frac{1}{2}}\left( \log N\right) ^{-1}\log c_{N}^{-1}$. On
the one hand, we have

\begin{equation}
\left(N^{-1}\log \log N\right)^{\frac{1}{2}} \log a_{N}^{-1} \sim \left(
N^{-1}\log \log N\right)^{\frac{1}{2}} \log c_{N}^{-1} + N^{-\frac{1}{2}} \left(\log \log N\right)^{\frac{1}{2}}\log N, \tag{A}
\end{equation}

\begin{equation*}
\frac{\left(\log \log n\right)^{\frac{1}{2}}}{c_{N}\log N}\left( \log
a_{N}^{-1}\right) ^{-1} \sim \frac{\left( \log \log N\right) ^{\frac{1}{2}}%
}{\left( \log c_{N}^{-1}\right) c_{N}\log N+cN\left( \log N\right)
^{2}\left( 1+o\left( 1\right) \right)} 
\end{equation*}

\begin{equation}
=\frac{\left( \log \log N\right) ^{\frac{1}{2}}}{\log N}\left( c_{N}\log
N\right) ^{-1}\left( 1+\frac{\log c_{N}^{-1}}{\log N}+o\left( 1\right)
\right) , \tag{B}
\end{equation}

\begin{equation}
\left( \left( \log \log n\right) ^{\frac{1}{2}}\left( a_{N}\log
a_{N}^{-1}\right) ^{\frac{1}{2}}\right) ^{2} \tag{C}
\end{equation}

\begin{equation}=c_{N}\left( \log
c_{N}^{-1}\right) N^{-1}\left( \log N\right) \left( \log \log N\right)
+c_{N}N^{-1}\left( \log N\right) ^{2}\left( \log \log n\right) \left(
1+o\left( 1\right) \right) .  \end{equation}

\noindent Obviously $(A)$, $(B)$ and $(C)$ together imply that the conditions of Part IV of Theorem \ref{tA}, namely,
as $N\rightarrow +\infty$,

\begin{equation}
c_{N}\rightarrow 0, \tag{W1}
\end{equation}

\begin{equation}
c_{N}\log N\rightarrow +\infty \tag{W2}
\end{equation}

\noindent and

\begin{equation}
\left( \log N\right) ^{-1}\left( \log c_{N}^{-1}\right) \left( \log \log
N\right) \rightarrow 0.  \tag{W3}
\end{equation}

\noindent In turn these facts imply the conditions (Q1-3-4). On the other hand, we have
\begin{equation*}
d_{N}\wedge _{N}\left( a_{N},\text{ }\bar{\alpha}_{N}\right) \leq
d_{N}\wedge _{N}\left( a_{N},\text{ }A_{N1}\right) 
\end{equation*}

\begin{equation*}
+0\left( \frac{\left(
a_{N}\log a_{N}^{-1}\right) \left( \log \log n\right) ^{\frac{1}{2}}N^{\frac{%
1}{2}}\log c_{N}^{-1}}{k^{\frac{1}{2}}\log N}\right) +0\left( \frac{\log
\log n}{\log N}\log c_{N}^{-1}\right) ,
\end{equation*}

\noindent a.s., as $N\rightarrow +\infty $, where we have used Lemma \ref{A2} and (\ref{3.1}%
). Further, as $N\rightarrow +\infty$,
\begin{equation}
\frac{\left( a_{N}\log a_{N}^{-1}\right) \left( \log \log n\right) ^{\frac{1%
}{2}}N^{\frac{1}{2}}\log c_{N}^{-1}}{k^{\frac{1}{2}}\log N}\sim \frac{\left(
\log a_{N}^{-1}\right) \left( \log \log N\right) \left( \log N\right) ^{3}}{%
\left( \log \text{ }N\right) \left( k^{\frac{1}{2}}\log N\right) N}%
c_{N}\rightarrow 0,  \tag{Q5}
\end{equation}

\noindent by the definition of $a_{N}$ and by $(W1)$ and $(W2)$. Thus, as $N\rightarrow
+\infty $, we have
\begin{equation*}
d_{N}\wedge _{N}\left( a_{N},\text{ }\bar{\alpha}_{N}\right) \leq
d_{N}\wedge _{N}\left( a_{N},\text{ }A_{N1}\right) +o\left( 1\right) ,a.s.
\end{equation*}

\noindent At this step, we apply Part II of Theorem 1 of \cite{7} by using Lemma \ref{A1}
which is true on account of $(Q1)$.\\

\noindent \textbf{Proof of Theorem \ref{tB}}.\\

\noindent We shall omit details of the proofs of the different parts that are the same
as those of the parts of Theorem \ref{tA}. The only problem concerns the bounds
depending on k. However, this problem is solved by Lemmas \ref{A3} and \ref{A4}. Hence we only provide the following remarks.\\

\noindent (R1) In our different choices of $\left( a_{N}\right) $, we have that $%
Na_{N}\rightarrow +\infty $, as $N\rightarrow +\infty $.\\

\noindent (R2) If $a_{N}\leq t_{k}\left( \delta \right) ,\delta >2$, then for any $y>0$%
, there exists $k_{y}$ such that 

\begin{equation*}
\forall k>k_{y}\text{, \ }k^{-y}N\leq \left( Nt_{k}\right) ^{-1}\leq \left(
Na_{N}\right) ^{-1}\rightarrow 0\text{ as }N\rightarrow +\infty.
\end{equation*}

\noindent (R3) $\left( \log \log n\right) =\left( \log \log N\right) \left( 1+o\left(
1\right) \right) $ and $\log n=\left( \log N\right) \left( 1+o\left(
1\right) \right) ,$ as $N\rightarrow +\infty.$\\

\noindent With these remarks, it is easily seen, as in the proof of Theorem \ref{tA} that the
conditions $(K)$, $(Q2)$, $(Q3)$, $(Q4)$ and $(Q5)$ are
satisfied at the same time with the specific assumptions of each part of
Theorem \ref{tB} as follows.\\

\noindent (a) $(Q2)$ and $(K)$ are always satisfied if $a_{N}=t_{k}.$ Indeed,
\begin{equation}
kN^{-1}\log \log n\sim \left( kN^{\frac{1}{2}}\right) \left( N^{-\frac{1}{2}%
}\log \log N\right) \rightarrow 0\text{ as }N\rightarrow +\infty,
\end{equation}

\noindent by $(R2)$ and $(K)$ and

\begin{equation}
k\left( \log \log n\right) \left( \log a_{N}^{-1}\right) \sim \left( N^{-%
\frac{1}{4}}k\log a_{N}^{-1}\right) \left( N^{-\frac{1}{4}}\log \log
N\right) \left( Na_{N}\right) ^{-\frac{1}{4}}\rightarrow,
\end{equation}

\noindent by $(R2)-(R3)$ and $(Q2$.\\

\bigskip

\noindent (b) In Parts I, II and III of Theorem \ref{tB}, the implication 
$\{(S1), (S3)\} \Rightarrow \{(Q3), (Q4)\}$ is true whenever $\log \log n\sim \log \log N$
(see the lines that follow Formula (\ref{3.9})) and $\left( Na_{N}\right)
\rightarrow +\infty $ as $N\rightarrow +\infty $, which are derived from
$(R1)$, $(R2)$ and $(R3)$.\\

\noindent (c) In Part IV, $(Q5)$ is true independently of the behavior of $k$.\\

\noindent Thus we may use Lemmas \ref{A3} and \ref{A4} instead of Lemmas \ref{A1} and \ref{A2} in the proofs
of Theorem \ref{tA} to get the results of Theorem \ref{tB} in the same way.\\

\newpage

\noindent \textbf{Proof of The Corollary}.\\

\noindent This is a direct consequence of Theorems \ref{tA} and \ref{tB} and of Lemma \ref{lem1}. For Part
III, the methods used in Part I of Theorem \ref{tA} must be repeated.\\

\begin{remark} \label{rD} By letting $\left( N\log \log N\right) ^{\frac{1}{4}}\left( \log
N\right) ^{\frac{1}{2}}\left( 2a_{N}\log a_{N}^{-1}\right) ^{-\frac{1}{2}%
}\rightarrow 0$, it would be possible to derive part I, III and IV of the
Theorem \ref{tA} from invariance principles such as in \cite{1} or \cite{5}. But
the necessary amount of work would be unchanged relatively to our method.
\end{remark}

{99}


\newpage
\begin{thebibliography}{99}
\Large

\bibitem{1} Aly, E.E.A., Beirlant, J. and Horv\`ath, L.(1984). Strong and weak
approximation of the k-spacings processes. \textit{Z. Wahrsch. verw. Gabiete}, 66,
461-484.

\bibitem{2} Cs\H{o}rg\"{o}, M. and R\'{e}v\`{e}sz, P. (1981). \textit{Strong Approximation in
Probability and Statistics}. Academic Press. New York.

\bibitem{3} Deheuvels, P. (1984). Spacings and applications. \textit{Techn. report 13, LSTA},
Universit\'{e} Paris 6.

\bibitem{4} K\'omlos, J. Major, M. and Tusn\`ady, G. (1975). Weak convergence and
embedding. In : \textit{Colloquia Math. Soc. Janos. Boylai. Limit theorems of
probability Theory}, 149-165. Amsterdam, North-Holland.

\bibitem{5} Lo, G.S. (1986). A strong upper bound in an improved approximation of
the empirical k-spacings process and strong limits for the oscillation
modulus. \textit{Techn. report 47, L.S.T.A.}, Universit\'{e} Paris 6.

\bibitem{6} Lo\`{e}ve, M. (1963). \textit{Probability Theory}. Van Nostrand Comp. Inc.
Princeton., 3rd ed.

\bibitem{7} Mason, D.M., Shorack, G. and Wellner, A. (1983). Strong limit theorems
for oscillation muduli of the empirical process. \textit{Z. Wahrsch. verw. Gabiete},
65, 83-87.

\bibitem{8} Pyke, R. (1972). Spacings. \textit{J. Roy. Statist. Soc}. Ser. B, 27, 359-449.

\bibitem{9} Shorack, G.R. (1972). Convergence of quantile and spacing process with
the application. Ann. \textit{Math. Statist}., 43, 1400-1411.

\bibitem{10}  Stute, W. (1982). The oscillation behavior of empirical process. \textit{Ann.
Probab}., 10, 86-107.
\end{thebibliography}
\end{document}